\documentclass{ws-ijmpd}
\newcommand{\co}{\cos\theta_1}
\newcommand{\si}{\sin\theta_1}
\newcommand{\cost}{\cos^{2}\theta_1}

\newcommand{\zo}{z_{1}}

\begin{document}

\markboth{Paolino and Pizzi}
{Electric force lines of the double Reissner-Nordstrom...}

%
\catchline{}{}{}{}{}
%

\title{Electric force lines of the double Reissner-Nordstrom exact solution}
\author{ARMANDO PAOLINO}

\address{Physics Department, University of Rome ``La Sapienza'', \\
Piazzale A. Moro, Rome, Italy 00185.
\\
armando.paolino@roma1.infn.it}

\author{MARCO PIZZI}

\address{Physics Department, University of Rome ``La Sapienza'',\\
Piazzale A. Moro, Rome, Italy 00185, and\\ ICRANet, Pescara, Italy 65122.
\\
pizzi@icra.it}

\maketitle

\begin{history}
\received{17-10-2007}
\revised{19-11-2007}
\comby{Managing Editor}
\end{history}

\maketitle

\begin{abstract}

 Recently, Alekseev and Belinski have presented a new exact solution of the Einstein-Maxwell equations which describes two Reissner-Nordstrom (RN) sources in reciprocal equilibrium (no struts nor strings); one source is a naked singularity, the other is a black hole: this is the only possible configuration for separable object, apart from the well-known extreme case ($m_i=e_i$). 
 
 In the present paper, after a brief summary of this solution, we study in some detail the coordinate systems used and the main features of the gravitational and electric fields. In particular we graph the plots of the electric force lines in three qualitatively different situations: equal-signed charges, opposite charges and the case of a naked singularity near a neutral black hole.
 
\end{abstract}

\section{Introduction}
 The new solution which has been recently found by Alekseev and Belinski\cite{AB} (in the following denoted with AB) has solved the long standing problem of the static equilibrium of two charged masses in the context of GR.
  
  While in the Newtonian theory the equilibrium condition is simply $m_1 m_2=e_1 e_2$, in the relativistic regime the problem is much more complicate because one has to solve the full system of the Einstein-Maxwell equations:
\begin{eqnarray}
	    \left\{
\begin{array}{l}
    R_{ij}-\frac{1}{2}R\,g_{ij}=2\left(F_{ik}F_{j}\,^{k}+\frac{1}{2}F_{lm}F^{lm}g_{ij}\right)\\\\
    (\sqrt{-g}F^{ik})_{,k}=\sqrt{-g}j^{i}
\end{array}
\right.
\end{eqnarray}
and find a static solution with two sources. Furthermore, in general this solution will present conic singularities at the symmetry axis\footnote{It is called ``conic singularity'' because the ratio between a small circumference around the axis and its radius is not $2\pi$ (as for a circle painted on a cone around its vertex).}; to find the equilibrium condition is equivalent to require \emph{the absence of any conic singularity}, i.e. the axis has to be locally Minkowskian ---physically that means that there must be neither ``struts'' nor ``strings''(see Ref.~\refcite{SS} for the rigorous relation between the value of the angle deficit and the effective energy-momentum tensor of these struts and strings) which prevent the two bodies to fall or run away each other.
 
 The key point to understand the main differences between classic and relativistic regime is the repulsive nature of gravity in GR near a naked singularity. This can be seen just by looking at the RN metric
\begin{equation}
 g_{tt}=1-\frac{2M}{r}+\frac{Q^2}{r^2},
\end{equation}
where gravity is repulsive for $r<\frac{Q^2}{M}$: it is for that reason that the equilibrium is allowed only at certain distances.
Indeed, e.g. if one considers the geodesic of a neutral particle on that background, it is easy to find a (stable) equilibrium precisely at
 \begin{equation}
 r_c=\frac{Q^2}{M}.
 \end{equation}
For charged particles an equilibrium is also possible at a fixed distance\cite{Bon}; in these cases it can be both stable or unstable, according to the choice of the parameters.
 
 In the AB-solution the Newtonian equilibrium condition is restored taking the limit of large distance between the two singularities.

 
  Although in principle such exact solution could be found already many years ago ---by using the Inverse Scattering Method (ISM) or the Integral Equation Method (IEM)---, practically nobody was able to eliminate the conic singularity  in a reasonable explicit way.  Indeed, the important achievement of the AB-solution is the extreme compactness of all the formulas, despite of complexity of calculations by which it was found\cite{AB2}. They get the wanted task using the IEM which presents some advantages with respect to the ISM\footnote{In the ISM there are also some unphysical parameters (NUT parameter, magnetic charge) and the rotation which are not easy to be eliminated.}.
 
 As they showed, the equilibrium is possible, apart from the well-known Majumdar-Papapetrou case where the charge of each source is equal to its mass, only for a naked singularity near a black hole (b.h.). We excluded from our analysis the b.h.-b.h. and naked-naked configurations since they do not exist at all in the equilibrium state.\\
 
 This paper is organized as follows: we give a brief historical review of the works in literature (Sec.~\ref{sec:History}) (this section can be skipped by the ones interested only to the physical contents); for the easy of the reader we also add the reproduction of the Alekseev-Belinski solution in Sec.~\ref{sec:Formulae}; we give some details clarifying the use of the coordinates systems involved (Sec.~\ref{sec:Detail}); then we recall the definition of the electric field in GR (Sec.~\ref{sec:FluxDefinition}) and finally we graph the plots of the electric force lines in the various qualitatively different cases (Sec.~\ref{sec:Plots}) ---which is the main task of our work.
 More precisely, in this last section, we consider at the beginning the general case with two charges, firstly with $e_1 e_2>0$ and then $e_1 e_2<0$; and finally that in which only one object (the naked singularity) is charged. The last particular case of the solution in different form was presented in Ref.~\refcite{AB3}. Of each case we present also the limit in which one source has a much smaller mass and charge than the other.
 
 In particular we consider the limit case of a small charged particle near a Schwarzschild black hole, finding electric force lines plots congruent with the Hanni-Ruffini\cite{HR} ones.

\section{Some Historical Remarks}
\label{sec:History}
The problem of the equilibrium of two charged masses and their resulting gravitational and electric fields has a long history in GR literature (see table \ref{tab:history}). It is possible to distinguish two different kind of results: approximate results, and exact solutions. 
 
 In the contest of the approximate results, the first to be mentioned is the one of Copson\cite{Cop}, who gave in 1927 the electric potential of a test charge on the Schwarzschild background (therefore it was neglected the backreaction of the particle on the metric tensor). That work was important because it gave the potential in a closed analytic form, however that result was not completely correct because it implicated that the black hole would have an induced charge: the correct potential was given by Linet\cite{Lin} only in 1976 ---the electric potential of the AB-solution indeed reduces to that form in the limit in which  the naked singularity source can be considered as a test particle. 

 In 1973 Hanni and Ruffini\cite{HR} gave for the first time the plots of the electric force lines\footnote{ We follow this work for the construction of the plots of the present solution.}, again for a test particle near a Schwarzschild black hole (but they used a multipole expansion of the electric potential).
 
 Later a certain number of papers have been published in which different authors (using exact solutions, PN and PPN approaches) arrived to different conclusions about the possibility/impossibility of an equilibrium configuration, however no final statements were achieved because of the use of supplementary hypothesis or for the incompleteness of the analysis. 
 
 In 1993 the already mentioned article of Bonnor\cite{Bon} gave an important hint to clarify the problem: studying the equilibrium configurations in the test particle limit, namely a test charge on the RN background, he pointed out that equilibrium configurations were possible when the ratio $e/m$ was less than unity for the background and more than unity for the particle, or \textsl{viceversa}; he showed also that equilibrium was possible for charges of opposite signs too. It is worth noting that the Alekseev-Belinski solution confirms practically word-by-word (from a qualitative point of view) that picture.
 
 Then in 1997 Perry and Cooperstock\cite{PC} found three numerical example showing that the equilibrium was possible for naked-b.h. configurations using an exact solution.

  Finally it is to mention the Bini-Geralico-Ruffini articles\cite{rr2,rr}, in which the authors found, using the Zerilli perturbative approach, the correction to the test particle approximation, considering the back-reaction of the particle to the background until the first order. Surprisingly they found that the Bonnor condition remain unchanged also considering these corrections.
 
 For what concerns the exact solutions history, the first two important articles were the ones of Majumdar and Papapetrou\cite{Maj,Pap}, which exhibited the fields of an arbitrary number of sources in reciprocal equilibrium, each one with $m_i=e_i$.
  
  For many years that was the only exact result known, the next step was made by Belinski and Zakharov\cite{BZ1,BZ2} in 1978  with the foundation of the Inverse Scattering Method in General Relativity (purely gravitational), which was then extended also to the Einstein-Maxwell equations by Alekseev\cite{Ale81} (see Ref.~\refcite{BV} for a self-consistent review). This method allows to find stationary, axially symmetric solutions with an arbitrary number of sources. From this time in principle the solution of our problem was available. However,  practically, the constraints necessary to eliminate the rotation, the conic singularity and the unphysical parameters (NUT parameter, magnetic charge) were too complicate to be handled analytically.

 The next step was made by Ernst and Hauser\cite{EH1,EH2}, Sibgatullin\cite{Sib} 
and Alekseev\cite{Ale85}, who developed different integral equation methods for 
constructing of solutions of Einstein-Maxwell equations. (The first 
method of such kind for pure gravity was already formulated in Ref.~\refcite{BZ1}). 
The method of Ref.~\refcite{Ale85} was used by Alekseev and Belinski to find the present 
solution\cite{AB} (see also Ref.~\refcite{AB2}), the important achievement of which is the 
extreme simplicity of the formulas and of the equilibrium condition.

\begin{table}
  \caption{Some historical remarks.}
  \label{tab:history}
  \begin{tabular}{cc}
    \hline
Perturbation Methods &  Exact Solutions \\
\hline

	 $\begin{array}{c}
	 Copson\ (1927)\\
	 \textbf{Electric field of a test charge}\\
	 \textbf{near a Schwarzschild b.h.}
	 \end{array} $ & \\\\
	 & $\begin{array}{c}
	 Majumdar-Papapetrou\ (1947)\\
	 m_i=e_i
	 \end{array}$

	 \\\\
$\begin{array}{c}
Hanni-Ruffini\ (1973)\\
$\textbf{Electric force lines of a test charge}$ \\ $ \textbf{near a Schwarzschild b.h.}$\end{array}$ & \\\\
$\begin{array}{c} 
     Linet\ (1976)\\
    $\textbf{A correction of Copson solution}$
     \end{array}$ & \\\\
                       & $\begin{array}{c} 
                     Belinski-Zakharov\ (1978) \\
                     $\textbf{Vacuum Solitons}$
                     \\$ and $ Alekseev\ (1980)\\
                     $\textbf{Electrovacuum Solitons}$
                        \end{array}$
              \\
             & $\begin{array}{c}
             $Solutions of $Hauser-Ernst\ (1979)\\ and\ Sibgatulling\ (1984)\, 
             \\$by IEM for rational axis data,$\\
             $and of $Alekseev (1985)\, 
             \\$by IEM for rational monodromy data$\\
             $\textbf{Integral Equation Method}$\end{array}$
             \\\\
$\begin{array}{c}Bonnor\ (1993)\\
  $\textbf{Equilibrium of a test particle}$\\
  $\textbf{on RN background}$
  \end{array} $ &\\\\
$\begin{array}{c}
             Perry-Cooperstock\ (1997) \\
             $\textbf{Equilibrium is possible}$\\
             $(3 numerical examples)$\end{array}$ &\\\\
$\begin{array}{c}
             Bini-Geralico-Ruffini\ (2007) \\
             $\textbf{Equilibrium of a test charge on RN}$\\
             $\textbf{with back-reaction until first order}$\end{array}$
                                           & $\begin{array}{c}Alekseev-Belinski\ (2007)\\
                                                 $\textbf{Exact solution for equilibrium}$\\
                                                 $ \textbf{(without strut) of two RN sources}$
                                               \end{array}    $          \\
\hline
  \end{tabular}
\end{table}

\section{Summary of the Alekseev-Belinski formulas}
\label{sec:Formulae}
The following (\ref{1.1})--(\ref{A.6}) formulas are the reproduction of formulas (1)-(10) of Ref.~\refcite{AB}.

 The solution, which can be interpreted as the non-linear superposition of two RN source at a fixed distance on the z-axis, is of the form
\begin{eqnarray}\label{1.1}
ds^{2}=H dt^{2}-\frac{\rho^{2}}{H}d\varphi^{2}-f(d\rho^{2}+dz^{2}) \\ \label{1.1b}
A_{t}=\Phi, \,\,   A_{\varphi}=A_{\rho}=A_{z}=0
\end{eqnarray}
where $H$ $f$ and $\Phi$ are real function of $\rho$ and $z$ only. In what follows $m_1,\ m_2$ and $e_1,\ e_2$ are the physical masses and charges of each source respectively\footnote{The expressions were found with the help of the Gauss theorem.}; the masses include also the interaction energy therefore $M_{tot}=m_{1}+m_{2}$, and $Q_{tot}=e_1+e_2$. 
It is convenient to use the spheroidal coordinates $(r_1,\theta_1)$ and $(r_2,\theta_2)$ which are linked to the Weyl coordinates $(\rho,z)$ by:
\begin{equation}\label{coo}
\begin{array}{l}
    \left\{
\begin{array}{l}
    \rho=\sqrt{(r_1-m_{1})^{2}-\sigma^2_{1}}\si
                       \\
                       z=z_{1}+\left(r_1-m_{1}\right)\co
\end{array}
\right.\\\\
 \left\{
\begin{array}{l}
     \rho=\sqrt{(r_{2}-m_{2})^{2}-\sigma^2_{2}}\sin\theta_{2}
                       \\z=z_{2}+\left(r_{2}-m_{2}\right)\cos\theta_{2}
\end{array}.
\right.\end{array}
\end{equation}
By definition $l\equiv z_2-z_1$ is the distance, expressed in the Weyl coordinate $z$, between the two objects.
Then, the explicit solution is:
\begin{eqnarray}\label{A.1}
H=&&\frac{[(r_1-m_1)^2-\sigma_1^2+\gamma^2\sin^2\theta_2][(r_2-m_2)^2-\sigma_2^2+\gamma^2\sin^2\theta_1]}{D^2}
\end{eqnarray}
\begin{eqnarray}\label{A.2}
\Phi =\frac{[(e_1-\gamma)(r_2-m_2)+(e_2+\gamma)(r_1-m_1)+\gamma(m_1\co +m_2\cos\theta_2)]}{D}
\end{eqnarray}
\begin{eqnarray}\label{A.3}
f=&&\frac{D^{2}}{[(r_1-m_1)^2-\sigma_1^2\cost][(r_2-m_2)^2-\sigma_2^2\cos^2\theta_2]}
\end{eqnarray}
where
\begin{eqnarray}\label{A.4}
D=r_1r_2-(e_1-\gamma-\gamma\cos\theta_2)(e_2+\gamma-\gamma\co),
\end{eqnarray}
while $\gamma$, $\sigma_1$ and $\sigma_2$ are defined by:
\begin{eqnarray}
\begin{array}{c}\label{A.5}
\gamma=(m_2e_1-m_1e_2)(l+m_1+m_2)^{-1}\,,\\  \\
 \sigma_1^2=m_1^2-e_1^2+2e_1\gamma\,, \ \ \ \ \sigma_2^2=m_2^2-e_2^2-2e_2\gamma.
 \end{array}
\end{eqnarray}

 It is easy to see that $(fH)_{\rho=0}=1$ on the whole axis, i.e. automatically there is no conic singularity.
 The above formulas give the solution satisfying the Einstein-Maxwell system only under the equilibrium condition
\begin{equation}\label{A.6}
m_1m_2=(e_1-\gamma)(e_2+\gamma).
\end{equation}

Each of the parameters $\sigma_1$ and $\sigma_2$ can be either real (in the case of a black hole) or imaginary (for a naked singularity); however in the following it will be always
\begin{equation}\label{1.2}
	\sigma^2_{1}>0 \,,\  \sigma^{2}_{2}<0,\ \ \texttt{and}\ \ \sigma_1>0\,
\end{equation}
i.e. the first source is ``dressed'' and the second is ``naked''. Since we want to deal only with separable objects, we require also the non-overcrossing condition 
 \begin{equation}\label{1.5}
	l-\sigma_1>0
\end{equation}
(it means that the naked singularity must be outside the horizon).
Using (\ref{A.6}), the distance $l$ can be written as a function of the other parameters by the very simple formula:
\begin{eqnarray}\label{1.3}
l=-m_1-m_2+\frac{m_1e_2-m_2e_1}{2(m_1 m_2-e_1 e_2)} \left[(e_2-e_1)\pm\sqrt{(e_1+e_2)^2-4\,m_1m_2}\right];
\end{eqnarray}
we always choose the sign in front of the root in (\ref{1.3}) in order to satisfy the non-overcrossing condition (\ref{1.5}).
From (\ref{1.3}) it is clear that the parameters must satisfy the restriction
 \begin{equation}\label{1.4}
	(e_2+e_1)^2>4\, m_1m_2.
\end{equation}


\section{Some further Details of the Solution}
\label{sec:Detail}

The solution has a very simple form, the only price to pay is just the simultaneous use of two pairs of coordinates.
Obviously for practical purposes, as for the electric lines plot, one needs the use of only one system ---in our case we choose $(r_1,\theta_1)$, the one related to the black hole (which is centered on the origin, since we took $\zo=0$ for simplicity, and consequently $z_2=l$). The linking relations are:
\begin{equation}\label{coo2}
\begin{array}{l}
\left\{\begin{array}{l}
    r_{2}-m_{2}=\frac{1}{\sqrt{2}}\sqrt{b^2+\sqrt{b^{4}-4\sigma^2_2(z-z_{2})^{2}}}
                       \\\\
    \cos\theta_{2}=(z-z_{2})(r_{2}-m_{2})^{-1}
\end{array},
\right.\end{array}
\end{equation}
where $b^2\equiv \rho^{2}+\sigma_2^2+(z-z_{2})^{2}$, while $\rho$ and $z$ have to be expressed using the first couple of (\ref{coo}). We take the plus sign of the roots in the first of (\ref{coo2}) since $r_1$ and $r_2$ must coincide at infinity.

 The peculiarity of the coordinates used needs a clarification in order to not misunderstand the physical properties of the solution, first of all where the ``true'' divergences are and what happens on the horizons.
\subsection*{Using $(r_1,\theta_1)$}
\label{sec:coord1}
These coordinates are centered on the black hole and can be considered as the natural generalization of the Schwarzschild ones.
For the peculiar choice of the $(r_1,\theta_1)-$coordinates, the horizon remains a perfect circle (it can be seen also analytically that $H$ vanishes at $r_h=m_{1}\pm \sigma_1$\, as for the single RN black hole). However the spherical symmetry is only apparent, indeed the invariants have a $\theta_1$-dependence and vary on the horizon.
 In this frame is not possible to reach the inside of the spheroid $r_2<m_2$ (we called the surface $r_2=m_2$ the `critical spheroid', as in \cite{AB}), therefore the second source (the naked RN centered in $z=z_2$), appears squeezed ``inside'' a horizontal segment that cuts the vertical axis: this happens because the naked singularity lies inside the region not covered by $(r_1,\theta_1)$.

\subsection*{Using $(r_2,\theta_2)$}
\label{sec:coord1}
Conversely, if one would to use $(r_2,\theta_2)$, the `critical spheroid' of the naked RN will appear as a sphere of coordinates $r_2=m_2$, while the black hole horizon as a segment squeezed on the axis: in this case it is the `critical spheroid' of the first source, i.e. $r_1<m_1$, that cannot be reached. Again, that has nothing to do with physics but just with the choice of the coordinate system).


 It is also to note the ``degeneracy'' of the Weyl coordinates: to the same point in $(\rho,z)$ it can corresponds different values of the spheroidal coordinates.


\subsection*{The electromagnetic invariant}
\label{sec:TheElectromagneticInvariant}
In order to understand where the charges are located it is useful to consider the electromagnetic invariant $\mathcal{F}=F^{ij}F_{ij}/2$. For the solution (\ref{1.1}) it has the form:
\begin{eqnarray}\label{F.1}
 \mathcal{F}& = & -\frac{[(r_1 - m_1)^2-\sigma_1^2](\partial_{r_1} \Phi)^2+(\partial_{\theta_1} \Phi)^2 }{f\,H\left[(r_1 - m_1)^2-\sigma_1^2\cos^2\theta_1\right]}.
\end{eqnarray}
It can be seen numerically (see tables~\ref{tab:bh},\ref{tab2}) that it diverges inside the horizon and \emph{inside} the critical spheroid of the naked RN \footnote{The spheroid, i.e. the line $\{0<\rho<\texttt{Im}(\sigma_1),\ z=z_2 \}$, seems apparently regular in $(r_1,\theta_1)$ coordinates just because its interior can be reached only using $(r_2,\theta_2)$}.
 
 It is also worth noting that on the critical spheroid, although in the $(r_1,\theta_1)$ representation it is a line, the up- and down-limit of $\mathcal{F}$ do not coincide, since they correspond to different points of the physical space-time.

Looking at $\mathcal{F}$ it is possible to see that no real discontinuity exists on the horizon, indeed it diverges only on the central singularities.
 
 The other invariant, $\epsilon^{ijkl}F_{ij}F_{kl}=\textbf{E}\textbf{B}$, is identically zero.


\section{Electric force lines definition}
\label{sec:FluxDefinition}
Just to understand better the meaning of the plots, we want to recall the definition of the electrical vector (which is not a trivial choice in GR). Following \cite{HR}, we define the electric field as the three non-diagonal time-like components of the controvariant tensor $F^{ij}$:
\begin{equation}\label{4}
\begin{array}{l l}
	 E^{\alpha}=F^{\alpha 0}\,, & \alpha=1,2,3.
\end{array}
\end{equation}
That identification is geometrically justified by the Gauss theorem generalized to the curved manifolds \cite{Whe}:
\begin{equation}\label{3}
4\pi Q=\int_{C}\textbf{*F}=\int_{C}\,^{*}F_{ij}dx^{i}\wedge dx^{j},
\end{equation}
where $\,^{*}F_{ij}=1/2\epsilon_{ijkl}F^{kl}\sqrt{-g}$ is the dual tensor of $F^{ij}$. Then it is natural to define the force lines in the usual way as the trajectoris of the dynamical system:
\begin{equation}\label{6}
\begin{array}{l} \left\{\begin{array}{c}
    \frac{d}{d\lambda}r_1 =E^{r_1} \\\\
    \frac{d}{d\lambda}\theta_1 =E^{\theta_1}
\end{array}
\right.\end{array}
\end{equation}
or equivalently by
\begin{equation}\label{fl.1}
  \frac{dr_1}{d\theta_1}=\frac{E^{r_1}}{E^{\theta_1}},\ \ \ \ \ \ \frac{E^{r_1}}{E^{\theta_1}}=((r_1-m_{1})^{2}-\sigma_{1}^{2})\frac{\partial_{r_1} \Phi}{\partial_{\theta_1} \Phi}.
\end{equation}
Then, from the equation of motion for this problem, restricting to our case, we have:
\begin{equation}\label{4a}
F^{r_1}\,_{t}u^{t}d\theta_1-F^{\theta_1}\,_{t}u^{t}dr_1=0,
\end{equation}
having used the coordinates $x^{i}=(t,\varphi,r_1,\theta_1)$.

The physical interpretation (Christodoulou-Ruffini, quoted in \cite{HR}) is the following: a force line is a line tangent to the direction of the electric force measured by a free-falling test charge momentarily at rest, with initial 4-velocity 
\begin{equation}\label{4ab}
u^{t}=(\sqrt{g^{tt}},0,0,0).
\end{equation}
Note that such interpretation is valid only for $g^{tt}>0$, for this reason we have not plotted the lines inside the horizon.

 In the $(t,\varphi,r_1,\theta_1)$ coordinates the metric (\ref{1.1}) becomes
\begin{equation}\label{4b}
\begin{array}{l}
	 ds^{2}=H dt^{2}-\frac{\rho^{2}}{H}d\varphi^{2}	              -f\left[(r_1-m_{1})^{2}-\sigma_{1}^{2}\cost\right]\left[\frac{dr_1^{2}}{(r_1-m_{1})^{2}-\sigma_{1}^{2}}+d\theta_1^{2}\right]
\end{array}
\end{equation}
while the electric potential remains unchanged. Then for the electric field we have:
\begin{equation}\label{5}
\begin{array}{l}
	 E^{\varphi}=0
	\\
	 E^{r_1}=g^{tt}g^{r_1r_1}\frac{\partial A_{t}}{\partial r_1}
	\\
	 E^{\theta_1}=g^{tt}g^{\theta_1\theta_1}\frac{\partial A_{t}}{\partial \theta_1}
\end{array}.
\end{equation}
Therefore the force lines are given by the solution of
\begin{equation}\label{fl.2}
  \frac{dr_1}{d\theta_1}=((r_1-m_{1})^{2}-\sigma_{1}^{2})\frac{\partial_{r_1} A_{t}}{\partial_{\theta_1} A_{t}}.
\end{equation}
It is worth noting that the force lines depend only on the two ratios $\partial_{r_1}A_{t}/\partial_{\theta_1}A_{t}$ and $g^{r_1r_1}/g^{\theta_1\theta_1}$ (indeed the conformal factor $f$ and neither $g_{tt}$ nor $g_{\varphi\varphi}$ do not appear in (\ref{fl.2})).
\section{Plots of the Force Lines}
\label{sec:Plots}

In the plots, what we called ``second source'' (i.e. the naked RN) is always \emph{up}, while the ``first source'' (i.e. the black hole) is always \emph{down} and centered on the origin.
 
 The lines are plotted in $(x,y)$ Cartesian coordinates defined as
\begin{equation}\label{fl3}
 \begin{array} {c}
 \left\{ \begin{array}{c}
            x=r_1\si \\
              y=r_1\co 
        \end{array}
\right.  \end{array}
\end{equation}
(they coincide with $(\rho,z)$ defined in (\ref{coo}) when $r_1\rightarrow\infty$).
\\
In the plots we have used geometrical units ($G=c=1$), in which the unitary length is given by the Schwarzschild mass $m_{1}=1$. 

 The graphical Faraday criterium is used, namely we plotted the electric force lines such that
\[
\frac{\textit{Number of lines from the first source}}{\textit{Number of lines from the second source}}\cong\frac{e_1}{e_2}.
\]

\subsection*{The separatrix}
\label{sec:TheSeparatrix.}
In general, when there are two charges, the electric force diagram will present a separatrix, which is a force line which reach asymptotically a saddle point of the potential and separates the lines of the two charges in the case they have the same sign, or ---in the case of opposite sign charge--- it delimits the region in which the lines flow from one to the other source. We marked these separatrix lines in bold; may be it is worth to mention that on the saddle point they have an invariant definition since on that point $\mathcal{F}=0$.

\subsection*{Inside the horizon}
\label{sec:InsideTheHorizon}

 In the following plots the force lines are graphed only outside the horizon since there it is no more possible to consider a static observer, i.e. the physical interpretation given in Sec. \ref{sec:FluxDefinition} does not hold because (\ref{4ab}) becomes imaginary). However, when the separatrix starts from the inside of the horizon, the study of that region is important to understand the difference between cases with the same or opposite charges. Therefore, in the case of fig. \ref{f.gen1B}, in which the saddle point is inside the horizon, we calculated the point where the separatrix touches the horizon, and we plotted the diagram just from there. (This was possible because mathematically the eqn. (\ref{fl.2}) is well defined also inside the horizon).
 

In the following three sub-sections we analyze the three qualitatively different sub-cases: $e_{1}e_{2}>0$ (\ref{sec:Equal}), $e_{1}e_{2}<0$ (\ref{sec:TwoChargesOfOppositeSign}), and finally $e_1=0$ (\ref{sec.One}).
\subsection{Two charges of equal sign ($e_1e_2>0$)}
\label{sec:Equal}
\subsubsection{\textbf{General case: two comparable RN sources}}
\label{sec:Comparable masses}
\begin{figure}
\centerline{\psfig{file=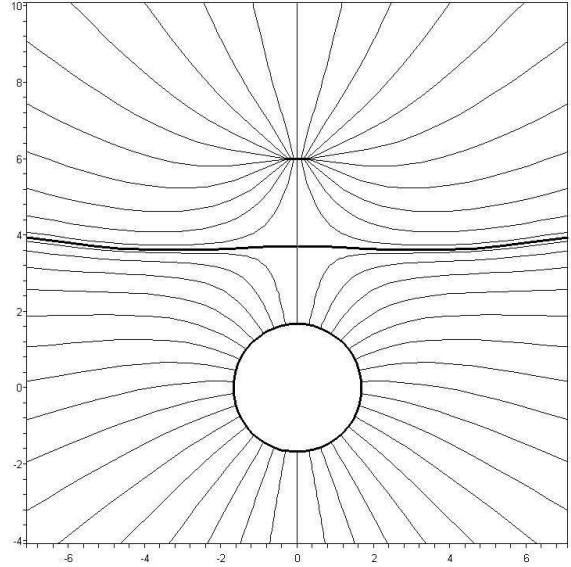,width=9cm}} 
\vspace*{8pt} 
  \caption{ Force lines in the general case (\ref{gen1A}), when the two RN have charges of the same sign. Note that the critical spheroid in that coordinate representation (\ref{fl3}) is an horizontal segment. The bold line is the separatrix. The circle on the bottom is the external horizon of the first source. Parameters used: $m_1=1$, $e_1=0.7$, $m_2=0.3$, $e_2=0.44$, $l=5$.}\label{RH-RN(A)-0}
\end{figure}

Let us consider the case in which the two RN sources have charges and masses of comparable dimensions
\begin{equation}\label{gen1A}
  \begin{array}{cc}
    m_{1} \approx m_{2} &    e_{1}\approx e_{2}  \\\\
    m_1^2>e_1^2 & m_2^2<e_2^2   \\\\
    e_1e_2>0.
  \end{array}
\end{equation}
This is the closest case to the classical picture, indeed here the equilibrium is mainly due to the classical balance of the electrostatic force and gravitational field. The resulting plot is given in fig.(\ref{RH-RN(A)-0}).

 The qualitative behavior of the force lines does not change with the changing of the distance $l$.
\subsubsection{Small charge (naked) near a RN black hole}
\label{sec:TestCharge}
The equilibrium configurations of this case with a small\footnote{ Here and in the following we say `small' charge and not `test' charge because the exact nature of the solution automatically takes in account all the back-reaction terms even if they can be very small (while the `test' limit is in general referred as the one in which all those terms are completely neglected).} charge (see fig. 2), namely with
\begin{equation}\label{Test1A}
  \begin{array}{cc}
    m_{1} >> m_{2}\,, &    |e_{1}|>> |e_{2}|  \\\\
    m_1^2>e_1^2 & m_2^2<e_2^2   \\\\
    e_1e_2>0\,,
  \end{array}
\end{equation}
have been studied in the test particle approximation first in Ref.~\refcite{Bon}, and recently in Ref.~\refcite{rr2}-\refcite{rr}, where they took in account also the back-reaction of the test particle.

\begin{figure}
\centerline{\psfig{file=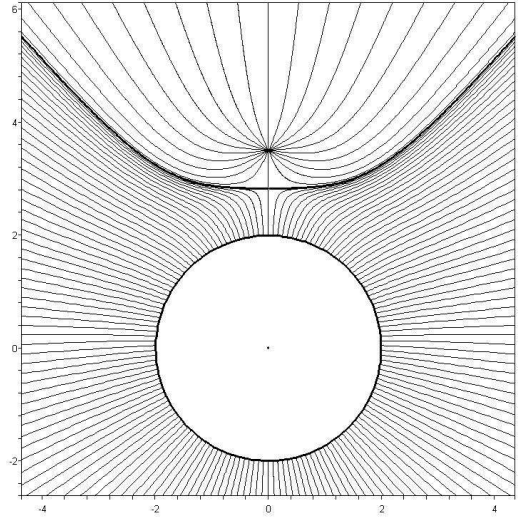,width=10cm}} 
\vspace*{8pt} 
  \caption{ Force lines of a small charge near a RN with horizon, case (\ref{Test1A}). Parameters used: $m_1=1$, $e_1=0.1$, $m_2=10^{-3}$, $e_2=1.3\cdot 10^{-2}$, $l=2.5$). The bold line is the separatrix.}\label{f.RH-TRN-0}
\end{figure}

\subsubsection{Small charge (with horizon) near a \textsl{naked} RN}
\label{sec:TestCharge}
This case does not exist for $e_1e_2>0$.

\subsection{Two charges of opposite sign ($e_1e_2<0$)}
\label{sec:TwoChargesOfOppositeSign}
 Although it is easy to show that in the previous cases with $e_1e_2>0$ the implications 
\begin{equation}\label{s1s2}
  \begin{array}{lcr}
    m_1^2>e_1^2 & \Rightarrow & \sigma_1^2>0
    \\\\
    m_2^2<e_2^2   & \Rightarrow & \sigma_2^2<0,
  \end{array}
\end{equation}
are always true, it is not so if $e_1e_2<0$. However in the following we considered two cases in which (\ref{s1s2}) holds.
\subsubsection{Two comparable RN sources}
\label{sec:Comparable masses}
\begin{figure}
\centerline{\psfig{file=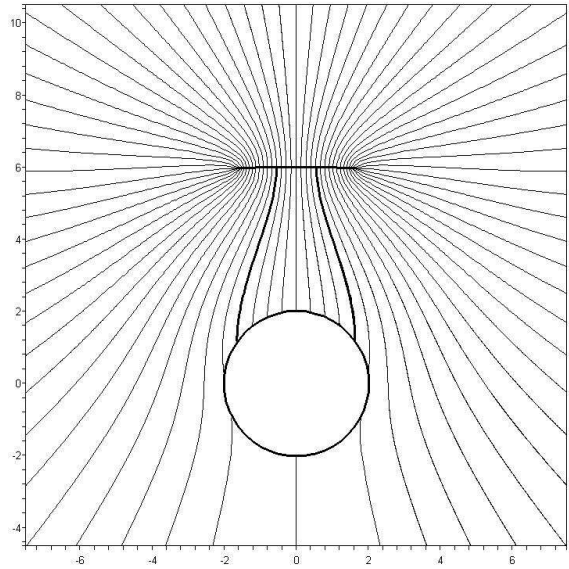,width=9cm}} 
\vspace*{8pt} 
  \caption{ Force lines in the general case (\ref{gen1B}), with charges of the opposite sign. Parameters used: $m_1=1$, $e_1=0.05$, $m_2=0.3$, $e_2=-1.66$, $l=5$.  The bold line is the separatrix, which now encircles also the central singularity of the b.h.: inside that region the lines go from one charge to the other. Outside that region the lines go from $e_2$ to infinity (some of them pass also through the horizon).}\label{f.gen1B}
\end{figure}

This case, with
\begin{equation}\label{gen1B}
  \begin{array}{cc}
    m_{1} \approx m_{2} &    e_{1}\approx -e_{2}  \\\\
    m_1^2>e_1^2 & m_2^2<e_2^2   \\\\
    e_1e_2<0,
  \end{array}
\end{equation} 
is the case in which the relativistic effects are much evident since here also the electric force is attractive (see fig. (\ref{f.gen1B})): in this case the equilibrium is due to the repulsive nature of the naked singularity.
\subsubsection{Small charge near a RN}
\label{sec:TestCharge}
It is also possible to find values that corresponds to a small charge with horizon near a naked RN:
\begin{equation}\label{Test1B}
  \begin{array}{cc}
    m_{1} << m_{2}\,, &    |e_{1}|<<|e_{2}|\,, \\ \\
    m_1^2>e_1^2\,, & m_2^2<e_2^2\,,   \\\\
    e_1e_2<0.
  \end{array}
\end{equation}
However in this case it would be useless to plot the force lines because the electric field is trivially Coulombian (the first source is weakly interacting both gravitationally and electrically).
\\\\
The inverse case, namely a small charge \emph{naked} near a RN \textsl{with horizon}, does not exist for particles lying outside the horizon (i.e. requiring $l>\sigma_1$), as noted by Bonnor\cite{Bon}.


\subsection{Cases with only one charge}
\label{sec.One}
In the following we will consider the cases with a naked singularity near a neutral black hole; they are qualitatively different from the previous ones since now there is no separatrix and the electric flux over the horizon surface is zero.

 In the particular case in which the first source is neutral (i.e. $e_{1}=0$), the equilibrium distance is even simpler,
\begin{equation}\label{1.6}	
l=-m_1-m_2+\frac{e_2^2}{2m_2}\left(1\pm\sqrt{1- 2\,m_1\left(\frac{e^2_2}{2m_2}\right)^{-1}}\right),
\end{equation}
which can be always satisfied for sufficiently large values of the charge parameter $e_{2}$.

\subsubsection{RN near a Schwarzschild black hole (comparable masses)}
\label{sec:RNNearSchwarzschildBlackHole}

\begin{figure}\label{f.S-RN-0}
 \centerline{\psfig{file=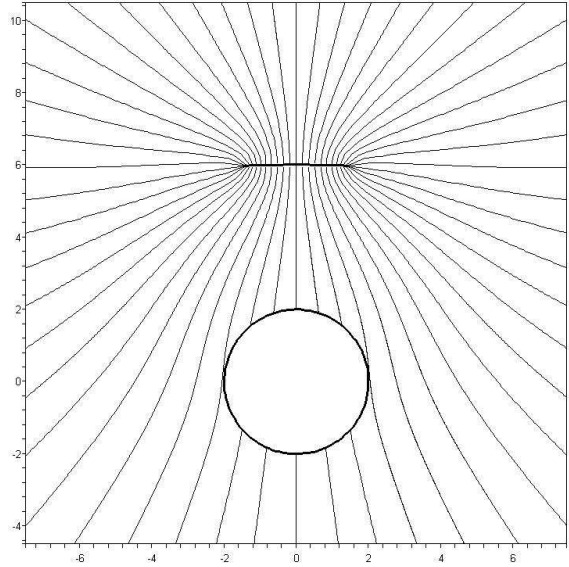,width=9cm}} 
\vspace*{8pt}
 \caption{ Force lines for the values (\ref{RNS}). The blank circle of radius $2m_{1}$ is the Schwarzschild horizon. Parameters used: $m_1=1$, $m_2=0.3$, $e_2=1.5$, $l=5$.}
\end{figure}

Thanks to the exact nature of the solution, it is very interesting also the case in which the RN source has comparable mass with the Schwarzschild black hole, say
\begin{equation}\label{RNS}
  \begin{array}{cc}
    m_{1} \approx m_{2}\,, &    e_{1}=0\,,  \\\\
    \sigma_1=m_1\,, & m_2^2<e_2^2\,,
  \end{array}
\end{equation}
indeed this case cannot be achieved by a perturbative approach, see fig. 4. It is possible to see that the electric lines are just slightly deformed by the gravitational field.

\subsubsection{Small charge near a Schwarzschild black hole}
\label{sec:HR}
We can also consider the small-charge limit, 
\begin{equation}\label{HR}
  \begin{array}{lr}
    m_{1} >> m_{2}\,,e_2 &    e_{1}=0 \,, \\\\
    \sigma_1=m_1 \,, & m_2^2<e_2^2\,,
  \end{array}
\end{equation}i.e. the second source is a small RN naked singularity. That is the only case in which we have a good comparing in literature, since it is the only case already studied (as we know) by using the force lines plots \cite{HR}, although by a perturbative approach. Strictly speaking the Hanni-Ruffini case refers to a slightly different situation, since they considered a particle \emph{momentarily} at rest in the Schwarzschild metric, while the AB solution is exactly static\footnote{From another point of view, Hanni-Ruffini do not use (\ref{1.6}) to determine the fourth parameter (because in their approximation the fourth parameter, say $m_2$, is considered arbitrarily small, therefore it is not present at all)}. However the present solution confirms very nearly their multipole expansion, since we find that the plots are in practice coincident.
In order to have the best possible comparing we considered the same distances between the charge and the horizon (figg. 5-7). Since now $l$ is not an independent parameter we fixed the masses values $m_{1}=1$ and $m_{2}=10^{-4}$, then varying the distance we found (using (\ref{1.3})) the relative parameter $e_{2}$. The test particle is at $z=l$, or equivalently at $r_1=l+m_1$. (Just to clarify the link with \cite{HR}'s notations: their $r$ is our $r_1$, and their $M$ is our $m_1$).

\begin{figure}
  \centerline{\psfig{file=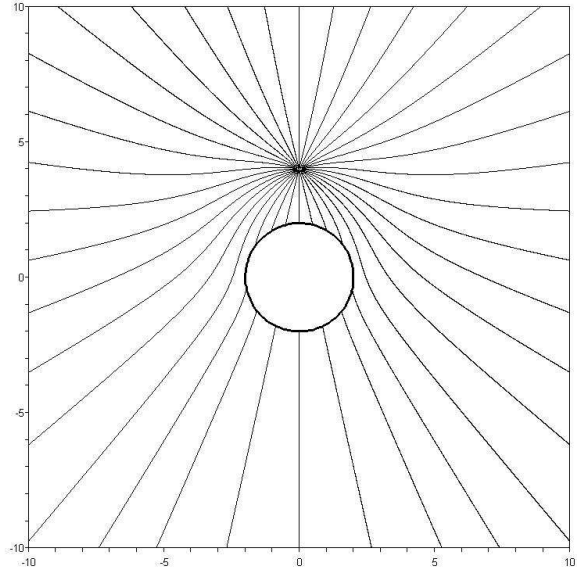,width=9cm}} 
  \caption{ Force lines for the values (\ref{HR}), with $l=3\,m_1$, i.e. in the spheroidal coordinates the particle is in $r_1=4\,m_1$. The circle of radius $2m_{1}$ is the Schwarzschild horizon. The plots are practically identical to the ones found by Hanni and Ruffini.}\label{f.HR1}
  \vspace*{8pt}
\end{figure}

\begin{figure}\label{f.HR2}
   \centerline{\psfig{file=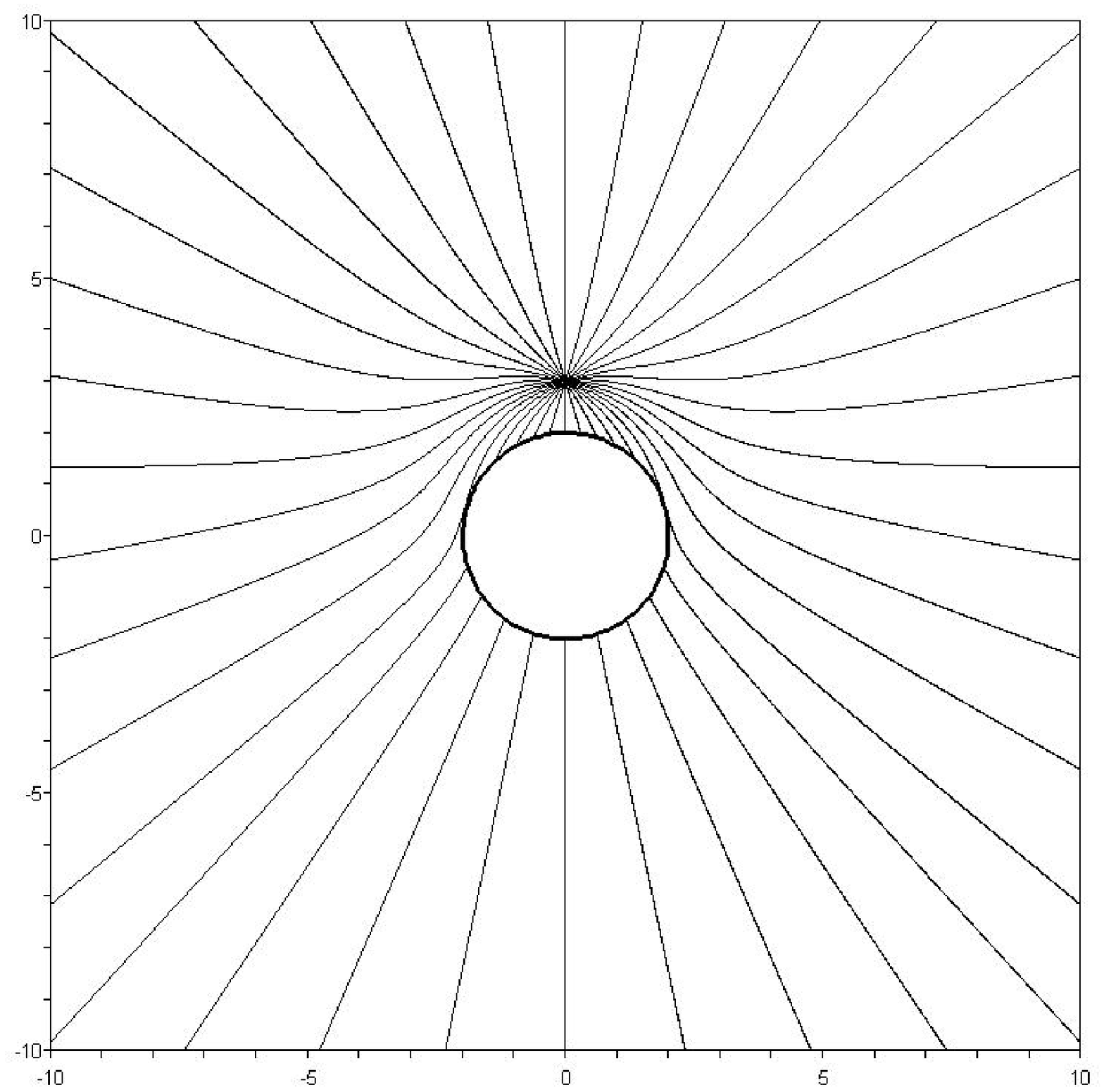,width=9cm}} 

 \caption{ Now the distance is $l=2\,m_1$, or equivalently the charge is in $r_1=3\,m_1$.}
\vspace*{8pt}
\end{figure}

\begin{figure}\label{f.HR3}
\centerline{\psfig{file=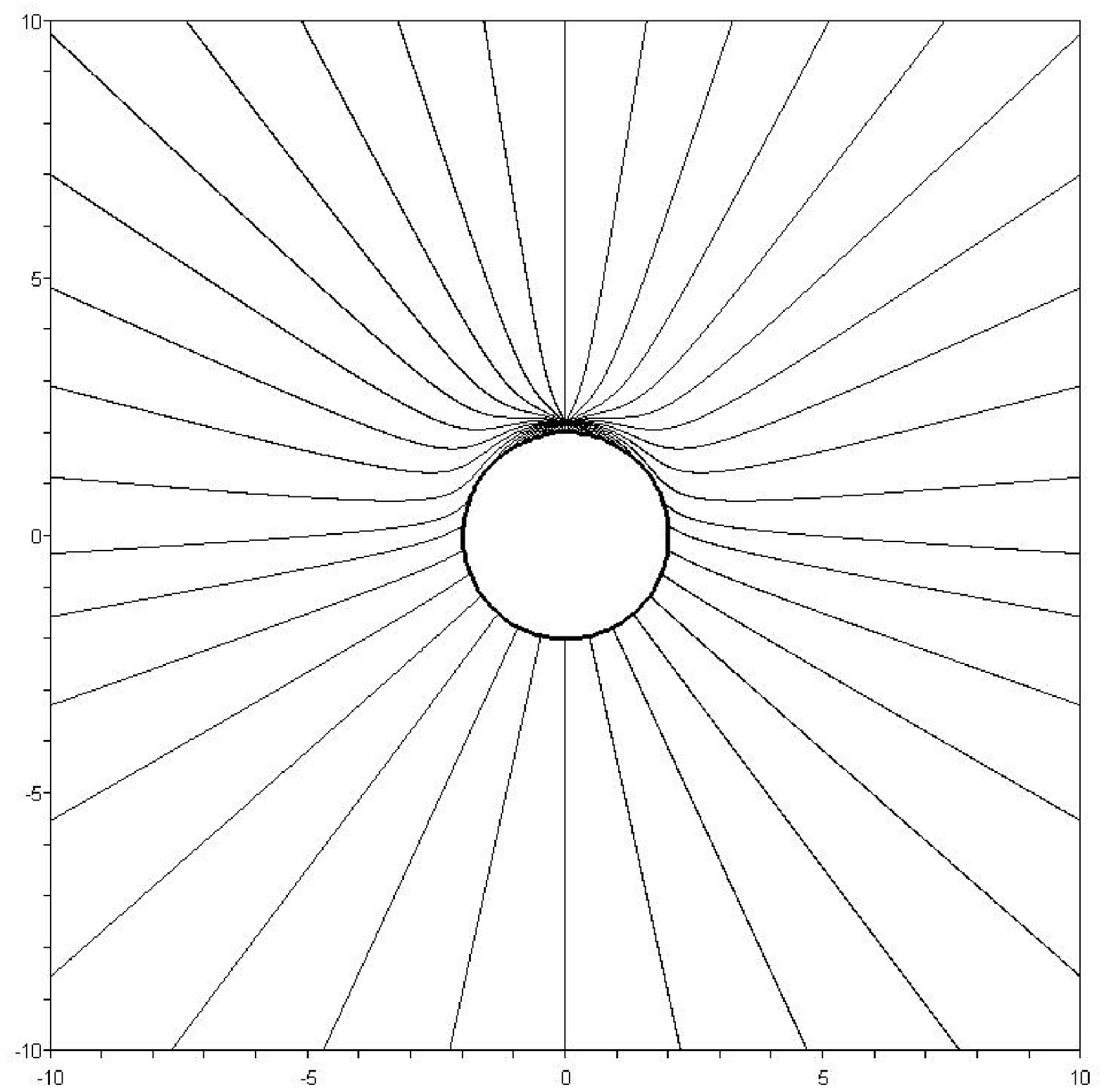,width=9cm}} 

 \caption{ Now the distance is $l=1.2\,m_1$, or equivalently the charge is in $r_1=2.2\,m_1$.}
\vspace*{8pt}
\end{figure}

From eq. (\ref{fl.1}), considering that now $\sigma_1=m_1$, it is easy to see that the corrections to the Hanni-Ruffini approximation are limited only on the exact form of the $A_{t}$ potential, since to use the Schwarzschild metric or the functions $H$ and $f$ given in (\ref{A.1}) and (\ref{A.3}) does not change the force lines.

\section{Final remarks}

 The main result of our analysis is that the exact solution seems to confirm quite strictly the test-charge approximation on the RN background (see. e.g. Ref.~\refcite{Bon}), which seems to give a good test of the exact picture.

\paragraph{Size of the naked singularity}
\label{sec:SizeOfTheNakedSingularity}
 Sometimes in literature  has been guessed (Ref.~\refcite{Fey}, cap.15; Ref.~\refcite{Gra}) that $e^2/2m$ should be considered as a `critical radius' of the naked singularity inside of which the RN solution has no physical meaning since it should be matched with a more realistic matter field tensor, in order to avoid the well known problems of a pointlike source, as the divergence of the electric energy. 
 
 If the quantity $e_2^2/2m_2$ can be roughly considered as the physical size of the RN charge, then from formula (\ref{1.6}) it is easy to see that the equilibrium configurations exist only for $e_2^2/2m_2$ larger than the Schwarzschild radius ($2m_1$). That seems to suggest that a real `small' charge limit cannot be achieved, in the sense that the particle can be `small' only gravitationally (and electrically), \emph{but not geometrically} because it would have a size larger than the Schwarzschild horizon. 
 
 However this is just a speculation since further investigations should be done to model the interior of a realistic RN source and find its radius.
  
\paragraph{Coordinate dependence of the plots}

Any plot of the force lines change drastically for different choices of the coordinates. However, what is interesting is to compare different situation by using the same coordinate representation, e.g. as we did for the Hanni-Ruffini case.

\paragraph{Stability}
If the solution would be unstable that would mean that it is a completely academic problem, since the equilibrium will be physically not allowed. However in the geodesic/test particle approximation ---which gives the essential features of the problem--- the equilibrium is stable, therefore at least in some range of values it should be the same also in the exact case (indeed the exact solution smoothly converge to the test particle approximation in the limit $e_i\,,\,m_i\rightarrow 0$, with $e_i/m_i$ finite, $i=1\ or\ 2$).
\\
\section*{Acknowledgments}
We wish to thanks V.A. Belinski and G.A. Alekseev for the valuable discussions.
One of us (M.P.) thanks also prof. Ruffini and ICRAnet for the financial support.

\bibliographystyle{unsrt}
\bibliography{bibExactForceLines}

\end{document}